\documentclass[runningheads,review]{llncs}
\usepackage[numbers,sort&compress]{natbib}

\usepackage{times}
\usepackage{latexsym}

\usepackage[T1]{fontenc}

\usepackage[utf8]{inputenc}

\usepackage{microtype}
\usepackage[hidelinks]{hyperref}
\usepackage{url}

\usepackage{amsfonts}
\usepackage{amssymb,amsmath}

\usepackage{mathtools}

\DeclarePairedDelimiterX{\infdivx}[2]{}{}{%
  #1\;\delimsize\|\;#2%
}

\usepackage{pifont}

\usepackage{graphicx}
\usepackage{booktabs}
\usepackage{diagbox}
\usepackage{multicol}
\usepackage{multirow}
\usepackage{todonotes}

\begin{document}
\title{Translate-Distill: Learning Cross-Language \\ Dense Retrieval by Translation and Distillation}
\titlerunning{Translate-Distill: Learning CL Dense Retrieval by Translation and Distillation}

\author{Eugene Yang\inst{1}\orcidID{0000-0002-0051-1535} \and 
Dawn Lawrie\inst{1}\orcidID{0000-0001-7347-7086} \and \\
James Mayfield\inst{1}\orcidID{0000-0003-3866-3013} \and
Douglas W. Oard\inst{1,2}\orcidID{0000-0002-1696-0407} \and
Scott Miller\inst{3}\orcidID{0009-0003-3345-6346}
}

\authorrunning{Yang et al.}
\institute{HLTCOE. Johns Hopkins University, Baltimore, MD 21211, USA \\
\email{\{eugene.yang,lawrie,mayfield\}@jhu.edu}
\and 
University of Maryland, College Park, MD 20742, USA \email{oard@umd.edu} 
\and
Information Sciences Institute, University of Southern California, CA 90292, USA \email{smiller@isi.edu}
}

\maketitle              %

\begin{abstract}
Prior work on English monolingual retrieval has shown that a cross-encoder trained using a large number of relevance judgments for query-document pairs can be used as a teacher to train more efficient, but similarly effective, dual-encoder student models.  Applying a similar knowledge distillation approach to training an efficient dual-encoder model for Cross-Language Information Retrieval (CLIR), where queries and documents are in different languages, is challenging due to the lack of a sufficiently large training collection when the query and document languages differ.  The state of the art for CLIR thus relies on translating queries, documents, or both from the large English MS MARCO training set, an approach called \textit{Translate-Train}. 
This paper proposes an alternative, \textit{Translate-Distill}, in which knowledge distillation from either a monolingual cross-encoder or a CLIR cross-encoder is used to train a dual-encoder CLIR student model. 
This richer design space enables the teacher model to perform inference in an optimized setting, while training the student model directly for CLIR.
Trained models and artifacts are publicly available on Huggingface.

\end{abstract}

\begin{figure}[t]
    \centering
    \includegraphics[width=\linewidth]{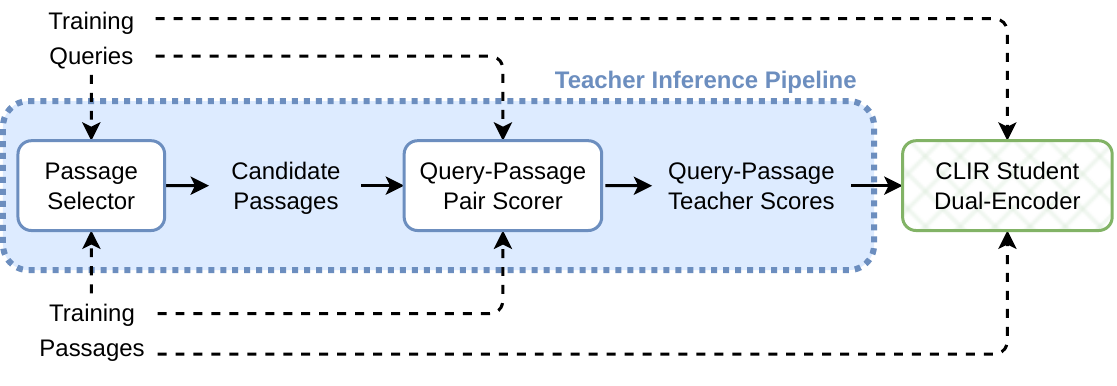}
    \caption{Translate-Distill training pipeline. The white boxes with blue borders are the fixed teacher models. The hatched green box is the trainable student model. Dashed arrows indicate the optional machine translation middle step, i.e., the input text the model receives can either be original or translated, with different translation decisions made for each dashed arrow.}
    \label{fig:flowchart}
\end{figure}

\section{Introduction}

Cross-language information retrieval (CLIR) enables users to express queries in one language and search for content in another.
Matching potentially ambiguous queries to documents is already challenging monolingually;
matching them across languages adds the additional complexity of translation. 
With recent improvements in machine translation, translating queries, documents, or both into the same language and then performing monolingual retrieval is a viable approach. 
However, machine translation is computationally expensive.
Query translation is particularly challenging because queries are often short,
writing styles for queries differ from the styles of documents
(for which machine translation systems are typically trained),
and tight query latency budgets provide little time for complex processing.
However, the alternative of translating very large document collections may not be feasible in cost-constrained applications. 
Therefore, this work---and a good deal of the work on CLIR---focuses on creating CLIR systems that do not require full neural machine translation during either indexing or query processing.

CLIR dual-encoders, such as DPR-X~\cite{c3} (also known as mDPR~\cite{mmarco}), ColBERT-X~\cite{colbertx} and BLADE~\cite{blade}, use multilingual pretrained language models 
(e.g., multilingual BERT~\cite{bert} or XLM-RoBERTa~\cite{xlmr})
to match queries to documents across languages
without using machine translation at indexing or retrieval time. 
However, the zero-shot approach of using English MS MARCO query-passage pairs to train a ColBERT dual-encoder using XLM-RoBERTa
has been shown suboptimal~\cite{colbertx}.
Instead, a more effective approach is to translate the MS MARCO pairs to match the CLIR setting of the final task before fine-tuning to the task,
an approach called \textit{Translate-Train}.  
This Translate-Train approach allows the model to simultaneously learn both the retrieval task and the details of the translation task for the specific language pair.
Translate-Train is the current state of the art for optimizing retrieval effectiveness in single-stage (i.e., ``end-to-end'') CLIR for large document collections~\cite{neuclir2022}. 

It is, however, possible to exceed the effectiveness of the single-stage approach by using cross-encoders as rerankers~\cite{lin2022pretrained, monobert}. 
CLIR cross-encoders, such as MonoT5 with mT5XXL~\cite{unicamp-at-neuclir}, are computationally expensive,
limiting their application in cost-constrained tasks. 
This paper addresses that limitation by using effective but inefficient cross-encoders at training time, as teacher models for training efficient CLIR dual-encoder models. 
Specifically, we propose \textit{Translate-Distill}
(summarized in Figure~\ref{fig:flowchart}),
an approach that distills ranking knowledge from cross-encoders through translations of the training data to create an effective CLIR dual-encoder model. 
While using distillation to train a monolingual dual-encoder retrieval model is not new~\cite{colbertv2,rocketqa}, generalizing the technique to train CLIR models is not straightforward,
both because the CLIR task has a larger design space for cross-encoder teacher models,
and because the choice of the teacher model can substantially influence the effectiveness of the resulting student model. 
In this work, we explore a suite of design options for Translate-Distill pipelines that use translation in different ways. 
Of particular importance are the languages used to express the query and the passages, which surprisingly do not need to be consistent throughout the training pipeline.

Our contribution is three-fold:
(1) introduction of a Translate-Distill training pipeline that distills knowledge from both cross-encoder and translation models; 
(2) a comprehensive analysis of the impact of each component in the Translate-Distill pipeline, with a recipe for training an effective CLIR dual-encoder model;
and (3) state-of-the-art CLIR dual-encoder models, benchmarked using the recent TREC 2022 NeuCLIR test collection. 
A Python implementation of the Translate-Distill pipeline is available on GitHub\footnote{\url{https://github.com/hltcoe/ColBERT-X/tree/plaid-x}.}, and the trained dual-encoder models and training materials (including the teacher models and scores) are available on Huggingface.\footnote{\url{https://huggingface.co/collections/hltcoe/translate-distill-659a11e0a7f2d2491780a6bb}.}

\section{Background}

Early work on CLIR explored the use of parallel corpora~\cite{landauer1991statistical}, comparable corpora~\cite{sheridan1996experiments}, and bilingual lexicons~\cite{pirkola1998effects} to cross the language barrier.
However, early work with shared semantic spaces using Cross-Language Latent Semantic Indexing (CL-LSI)~\cite{rehder1998automatic} was not widely adopted, largely for efficiency reasons.
Using full neural document translation to reduce CLIR to a monolingual task is now substantially more effective than all of these earlier techniques,
although translation costs can limit its use on large collections~\cite{blade}.
Query translation is typically less effective, however, as language use in queries differs in important ways from that in documents, and training resources that are specialized for query translation are not widely available. 
The research frontier is thus currently focused on balancing effectiveness and efficiency, seeking to develop approaches that are as effective as, and considerably more efficient at indexing time than, full neural machine translation~\cite{neuclir2022}.

Cross-encoders, such as MonoBERT~\cite{monobert} or MonoT5~\cite{monot5}, are a family of retrieval models that use neural pretrained language models (PLM) such as BERT~\cite{bert}, RoBERTa~\cite{roberta}, or T5~\cite{t5} by concatenating the query and document as a single input text sequence;
this allows cross-attention between the query and document representations. 
Although very effective, such models require that the query be available before each document can be processed.
This prevents document representations from being indexed in advance,
thus limiting the number of documents to which the technique can be applied.  
Cross-encoders are thus primarily used as rerankers in a retrieval pipeline,
following an efficient retrieval model such as BM25;
this is known as a \textit{Retrieve-and-Rerank} pipeline~\cite{monobert}. 

On the other hand, dual-encoders (e.g., DPR~\cite{dpr}, SPLADE~\cite{formal2021splade}, and ColBERT~\cite{colbert}) are a family of model architectures that encode queries and documents separately, using a PLM as the encoder. 
This is a modern neural realization of LSI's shared semantic space,
with the additional benefit of the non-linearity that neural models make possible
(LSI was limited to lossy linear transformations on the term-document matrix). 
These dual encoder models enable offline document preprocessing before any query has arrived,
thus (when used with efficient approximate nearest neighbor techniques) making it possible to respond to queries quickly, even for large collections.

English neural retrieval models use collections with millions of judged query-passage pairs,
such as MS MARCO~\cite{msmarco} and Natural Questions~\cite{kwiatkowski2019natural},
as training data. 
Until similar cross-language training data becomes available for a broad range of language pairs,
training neural CLIR models will remain more complicated than training monolingual models.
While it is possible to train a model that uses an mPLM with English query-passage pairs zero-shot,
prior work has found that providing cross-language supervision during training leads to more effective CLIR models~\cite{colbertx, c3, blade}.  
Such supervision can be in the form of pretraining,
such as by using XLM-Align~\cite{xlm-align} as the mPLM~\cite{c3},
continued pretraining of an mPLM with parallel or comparable text for a specific language pair~\cite{c3, blade},
or fine-tuning with queries and documents that have been translated into the languages of the final CLIR task~\cite{colbertx}. 
These training strategies strengthen the model's ability to bridge the query/document language barrier.

The top-ranked run for each document language in the TREC 2022 NeuCLIR track~\cite{neuclir2022}
used a cross-encoder reranker at the end of a retrieval pipeline~\cite{unicamp-at-neuclir}.
Each used MonoT5 with the mT5XXL mPLM,
trained from mMARCO~\cite{mmarco}
and document-language queries produced by machine translation
to rerank the top-ranked documents.
We refer to this reranker as \textit{Mono-mT5XXL}. 
Although effective,
Mono-mT5XXL has 13 billion parameters, requiring GPUs with large memory or special model-sharding approaches. 
To exploit the effectiveness of cross-encoder rerankers and the efficiency of dual-encoders,
this paper explores distilling ranking knowledge from an effective cross-encoder, such as Mono-mT5XXL, to train a CLIR dual-encoder model. 

Prior work on English retrieval has explored distillation for training monolingual retrieval models~\cite{rocketqa, colbertv2, formal2022distillation, lin2023prod, zeng2022curriculum}.
Such approaches use an English cross-encoder as a teacher model to score training query-passage pairs,
using those scores to optimize a student dual-encoder model with KL-divergence loss.
\citet{li2022learning} jointly trained a cross-language alignment model
(a form of translation model)
with a distillation objective from an English dual-encoder as teacher and a CLIR ColBERT model as student.
While effective, this training requires a parallel corpus with IR relevance judgments, which are rarer than unicorns. Therefore, they used XOR-TyDi~\cite{xortydi}, a synthetic multilingual retrieval collection built from Wikipedia, for supervision on alignment and ranking.
This is similar to prior work on continued pretraining with comparable text~\cite{c3, sun2020clirmatrix}. 
We explore an alternative approach.

Given that Translate-Train has been shown to be an effective training strategy
and that it only requires translation of the training corpora, such as MS MARCO,
we propose a unified framework \textit{Translate-Distill}.
Our framework combines Translate-Train with cross-encoder distillation to train a CLIR dual-encoder model without additional data resources. 
In the next section, we introduce the Translate-Distill framework, and we identify a series of design choices in a Translate-Distill pipeline.

\section{Translate-Distill}

Illustrated in Figure~\ref{fig:flowchart}, our proposed Translate-Distill pipeline consists of two inference steps using teacher models and a training step for the student CLIR dual-encoder model using knowledge distillation. 
We first introduce the distillation training framework for CLIR and then discuss the pipeline for generating the training material. 
Of particular importance is the languages of the queries and passages, which is not necessarily consistent throughout the pipeline. 

\subsection{Notation}

\newcommand{\lang}{\mathcal{L}}
\newcommand{\qlang}{\mathcal{L}_\mathbb{Q}}
\newcommand{\dlang}{\mathcal{L}_\mathbb{D}}
\newcommand{\tlang}{\mathcal{L}_T}
\newcommand{\trans}[1]{\widetilde{#1}}

Let $\qlang$ and $\dlang$ be the query and the document languages of the final CLIR task.
In this work, we focus on CLIR passage retrieval, limiting the retrieved items to a fixed length,
due to the limitations of current transformer models.
We do, however, map those passages back to the documents they came from
when evaluating using the NeuCLIR 2022 test collection.
We follow the convention in the CLIR literature of using the name ``document language'' to refer to the language of the passages we retrieve. 
We assume that the original training collection is monolingual, containing a set of training queries $\mathcal{Q}$ and passages $\mathcal{P}$ in the same language $\tlang$. 

We use $\trans{\lang}$ to denote machine translation into language $\lang$.
We place this notation as a superscript to indicate the language into which text has been translated.
For example, machine-translated queries that have been translated into the document language $\dlang$ are denoted as $\mathcal{Q}^{\trans{\dlang}}$.
We also use this notation in compound superscripts to denote the languages on which functions operate.
For example, a function denoted $f^{\lang_a, \lang_b}$ accepts a pair of texts in language $\lang_a$ and $\lang_b$.
For convenience, we define $\mathcal{Q}^{\trans{\tlang}} = \mathcal{Q}^{\tlang}$ and $\mathcal{D}^{\trans{\tlang}} = \mathcal{D}^{\tlang}$ since the text in $\mathcal{Q}$ and $\mathcal{P}$ are already written in language $\tlang$.

\subsection{Cross-language Knowledge Distillation}

To select the training passages for each query $q$, we
retrieve the top $k$ passages from $\mathcal{P}$.
Specifically, given a passage selector\footnote{What we call a passage selector assigns scores to passages that are the basis for selection.  We call it a selector rather than a scorer to avoid confusion with the query-passage scorer.} accepting queries in language $\lang_a$ and passages in language $\lang_b$ as $PS^{\lang_a, \lang_b}(\cdot) \rightarrow \mathbb{R}$,
we define the set of candidate passages $\mathcal{C}_q \subset P$ of size $k$ for query $q\in\mathcal{Q}$ such that 
\begin{align*}
    PS^{\lang_a, \lang_b}(q^{\widetilde{\lang_a}}, p^{\widetilde{\lang_b}}_{in}) \ge PS^{(\lang_a, \lang_b)}(q^{\widetilde{\lang_a}}, p^{\widetilde{\lang_b}}_{out}) 
    \quad \forall p_{in} \in \mathcal{C}_q \text{ and }  p_{out} \notin \mathcal{C}_q
\end{align*}
Passages in $\mathcal{C}_q$ are similar to the ``hard negatives'' for training IR models. 
However, in our approach, these candidate passages may be relevant to the queries. 
$\mathcal{C}_q$ forms the ``teaching material'' for the subsequent CLIR student dual-encoder model. 

For a trainable CLIR student dual-encoder model $DE^{\qlang, \dlang}(\cdot) \rightarrow \mathbb{R}$,
we construct a distillation loss using KL-divergence against a fixed query-passage pair scorer $QP^{\lang_c, \lang_d}(\cdot) \rightarrow \mathbb{R}$,
typically an effective cross-encoder model.
Specifically, we define the loss function for Translate-Distill $L_{TD}$ based on the predicted query-passage scores from the teacher and student as 
\begin{align}
    L_{TD}(q) &= D_{KL}\left(  DE^{\qlang, \dlang} \;||\; QP^{\lang_c, \lang_d} \right)  \nonumber \\
         &= \sum_{p \in \mathcal{C}_q} DE^{\qlang, \dlang}(q^{\trans{\qlang}}, p^{\trans{\dlang}}) 
           \log\left( \frac{
                DE^{\qlang, \dlang}(q^{\trans{\qlang}}, p^{\trans{\dlang}})
            }{
                QP^{\lang_c, \lang_d}(q^{\trans{\lang_c}}, p^{\trans{\lang_d}})
            } \right) \label{eq:td-loss}
\end{align}

The pair scorer $QP^{\lang_c, \lang_d}$ provides supervision to the student dual-encoder $DE^{\qlang, \dlang}$ based on $\mathcal{C}_q$. 
Therefore, if the $\mathcal{C}_q$ is not informative or sufficiently hard, the cross-encoder cannot adequately distill its ranking knowledge. 

Constructing the training loss by Equation~(\ref{eq:td-loss}) has the important benefit of decoupling the input languages of the three modules in the pipeline. 
We can optimize the effectiveness of each teacher model by choosing languages that matches its training,
avoiding language mismatches between training and inference time on a module-by-module basis. 
Similarly, we can choose the languages of the input text for the student model based on how we intend to use that model at inference time. 
Such decoupling avoids unnecessary language transfer, and enables each module to operate or train in the languages for which it was designed. 

The input languages for the teacher model's passage selector $PS$ ($\lang_a$ and $\lang_b$) and query-passage pair scorer $QP$ ($\lang_c$, and $\lang_d$) could also be different. 
However, in practice, they are often chosen to be among $\qlang$, $\dlang$, and $\tlang$ to align with either the final CLIR task or the available monolingual training resource.
In the rest of this section, we discuss language selection given varying constraints. 

\subsection{Training Pipeline}

\begin{table}[t]
\setlength\tabcolsep{0.4em}
\caption{Example input text query-passage language pairs for each module in the Translate-Distill Pipeline. These examples assume that the final CLIR task requires English (Eng) queries and Persian (Fas) documents,
and the training resources are all in English. %
}
\centering
\begin{tabular}{l|cc|c|c}
\toprule
                       &  \multicolumn{2}{c|}{Teacher Inference} &  Training & Retrieval \\
\midrule
                                         &  Passage  & Query-Passage  & CLIR Student    & CLIR Student  \\ 
Availability of Key Modules              &  Selector & Pair Scorer    & Dual-encoder    & Dual-encoder  \\ 
\midrule
Eng Training Query \& Passages           &  Eng-Eng  &  Eng-Eng  &  Eng-Eng  &  Eng-Fas \\
+ Eng-Fas MT model                       &  Eng-Eng  &  Eng-Eng  &  Eng-Fas  &  Eng-Fas \\
\hspace{2mm} + CLIR cross-encoder        &  Eng-Eng  &  Eng-Fas  &  Eng-Fas  &  Eng-Fas \\  
\hspace{2mm} + Multilingual Cross-encoder&  Eng-Eng  &  Fas-Fas  &  Eng-Fas  &  Eng-Fas \\
\hspace{4mm} + E2E CLIR Retrieval System &  Eng-Fas  &  Fas-Fas  &  Eng-Fas  &  Eng-Fas \\
\bottomrule
\end{tabular}    
\label{tab:example-lang-pair}
\end{table}

The Translate-Distill pipeline is illustrated in Figure~\ref{fig:flowchart}.
We pre-construct (1) the translations of $\mathcal{Q}$ and $\mathcal{P}$;
(2) the candidate passages $\mathcal{C}_q$ for each query $q$;
and (3) the teacher scores $s_{q,p} = QP^{\lang_c, \lang_d}(q^{\trans{\lang_c}, p^{\trans{\lang_d}}})$ before training. 
Although all these resources can be generated on-the-fly during the dual-encoder training,
pre-computing them limits peak computational resource requirements during training.
Training queries and passages are first fed into the passage selector to select $\mathcal{C}_q$ in the language pairs that the selector accepts.
For each $p\in\mathcal{C}_q$, we use the query-passage pair scorer to produce the teacher score for each $(q, p)$ pair.
These scores are stored and later loaded back into memory when training the CLIR student dual-encoder model.  

The first row in Table~\ref{tab:example-lang-pair}
is an English training setup;
the remaining four rows illustrate a series of possible input language choices of the teacher and student models for an English-Persian CLIR task. 
The Translate-Distill training pipeline requires the availability of machine translation models
that can translate the training queries and passages from $\tlang$ to $\qlang$ and $\dlang$.
More precisely, it requires the existence of $\mathcal{Q}^{\trans{\qlang}}$ and $\mathcal{D}^{\trans{\dlang}}$.
For example, the publicly available neuMARCO~\cite{neuclir2022} and mMARCO~\cite{mmarco} collections provide machine translation of the popular training collection MS MARCO~\cite{msmarco}. 
Such requirements are identical to those of Translate-Train. 

If effective cross-encoders for the document language are available,
either using CLIR or using queries in that same language,
we can use teacher models that more closely match the final CLIR setting. 
When the original language of the training passages is English rather than Persian,
a CLIR or a multilingual cross-encoder could directly produce teacher scores using translated Persian passages. 
To see why this could be helpful, consider the case of a passage that is relevant in its original English version, but that becomes non-relevant after translation due to translation errors.
A teacher cross-encoder that directly (and correctly) scores Persian passages would reflect such shifts in its scores. 
This approach reduces the mismatch of the input text between the teacher and student models. 

Similarly, if a CLIR system already exists, we could use this CLIR system to select the candidate passages $\mathcal{C}_q$ for each query $q$.
This approach would provide in-language hard negatives,
helping the student model learn to distinguish relevant and non-relevant passages as they are expressed in the document language. 

Of course, the benefits to be gained from such approaches depend on the quality of the multilingual or CLIR systems used for those purposes;
Section~\ref{sec:results} demonstrates how using them can improve distillation results.

\section{Experiments}

This section introduces our evaluation collections and metrics,
and discusses resources and models used in our Translate-Distill pipeline experiments. 

\subsection{Evaluation Collections and Metrics}

\begin{table}[t]
\setlength\tabcolsep{0.8em}
\caption{Collection statistics. }
\centering

\begin{tabular}{l|rrr|rrr}
\toprule
Collection    &  \multicolumn{3}{c|}{NeuCLIR 2022} & \multicolumn{3}{c}{HC3} \\ %
Language      &  
\multicolumn{1}{c}{zho} & \multicolumn{1}{c}{fas} & \multicolumn{1}{c|}{rus} &
\multicolumn{1}{c}{zho} & \multicolumn{1}{c}{fas} & \multicolumn{1}{c}{rus} \\ %
\midrule
\# Documents  &  3.2M &  2.2M &  4.6M &  5.6M &  7.3M & 26.8M \\ %
\# Passages   & 19.8M & 14.0M & 25.1M &  6.3M &  0.4M & 27.0M \\ %
\midrule
\# Topics     &    49 &    46 &    45 &    50 &    50 &    69 \\ %
\bottomrule
\end{tabular}

\label{tab:collection-stats}
\end{table}

We evaluate Translate-Distill on the TREC 2022 NeuCLIR track~\cite{neuclir2022} and HC3~\cite{hc3} test collections. 
Collection statistics are summarized in Table~\ref{tab:collection-stats}.
Both of these test collections model three CLIR tasks:
searching Chinese, Persian, or Russian documents using English queries. 
The NeuCLIR collections consist of news articles, a genre for which many machine translation models are optimized, extracted from Common Crawl.
HC3, by contrast, consists of short informal Twitter conversations,
which can pose challenges for machine translation. 
We report nDCG@20 (the primary evaluation metric for NeuCLIR 2022) as our effectiveness score. 
We use the topic title concatenated with the description as the query.
Note that prior work has found that reversing the order of the title and description in the query
(i.e., description followed by title) is more effective for Chinese~\cite{unicamp-at-neuclir}.
However, for experimental consistency we retain the conventional order for all languages and collections.  

\subsection{Training Queries and Passages}

We use MS MARCOv1 training queries and passages~\cite{msmarco} as our training set. 
Passage translations were provided by the NeuCLIR organizers and released along with the test collection with the name \textit{neuMARCO}~\cite{neuclir2022} on \texttt{ir-datasets}~\cite{irdatasets}.\footnote{\url{https://ir-datasets.com/neumarco.html}} 
These translations are generated by Sockeye v2 trained with general domain parallel text.
MS MARCO training query translations were obtained from mMARCO~\cite{mmarco}.
These translations are generated using the Google Translate service. 
Since Persian is not included in mMARCO, we used Google Translate ourselves to generate Persian translations of the queries.\footnote{\url{https://huggingface.co/datasets/hltcoe/tdist-msmarco-scores/blob/main/msmarco.train.query.fas.tsv.gz}} 
For consistency, although mMARCO also provides translation for Chinese and Russian passages, we use the neuMARCO version in our experiments whenever passage translation is needed. 

\subsection{Teacher and Student Models}
We use publicly-available models as passage selector and teacher query-passage pair scorer. 
In the Translate-Distill pipeline, we select the top 50 passages from the MS MARCOv1 passage collection for each training query. 

\subsubsection{Passage Selector}
For our main experiments in Section~\ref{sec:main} and Table~\ref{tab:main} we use the ColBERTv2~\cite{colbertv2} model released by the ColBERT authors as the passage selector. 
This is a monolingual English retrieval model;
thus, the input query and passages are not translated, but kept in English. 
To assess training robustness, we also experiment using different retrieval systems as passage selectors in contrastive experiments. 
As one alternative, for English, we experiment with CoCondenser~\cite{cocondenser},  a single-vector dual-encoder model.\footnote{We use the hard negatives from \texttt{co-condenser-margin\_mse-sym\_mnrl-mean-v1}, which are publicly available on Huggingface Datasets released by the Sentence-Transformers: \url{https://huggingface.co/datasets/sentence-transformers/msmarco-hard-negatives}. }
Results using these alternatives appear in Table~\ref{tab:diff-selector}.
Assuming the existence of an end-to-end CLIR system,
we experiment with using an existing Translate-Train ColBERT-X model~\cite{colbertx} to select candidate passages. 

\subsubsection{Query-Passage Pair Scorers.}
We experimented with MiniLM~\cite{minilm}, MonoT5-3b~\cite{monot5}, Mono-mT5XXL~\cite{unicamp-at-neuclir}, and four
cross-encoders
(one English-Trained and three Translate-Trained models on Chinese, Persian, and Russian passages)
that we trained ourselves. 
MiniLM (a lightweight model distilled from a large BERT cross-encoder) and MonoT5-3b are commonly used cross-encoder rerankers that have very different resource requirements;
Mono-mT5XXL is the state-of-the-art cross-encoder for CLIR
(according to the NeuCLIR 2022 evaluation results~\cite{neuclir2022}). 
MonoT5-3b and Mono-mT5XXL are executed on 2 NVidia 40GB A100 GPUs with DeepSpeed~\cite{rasley2020deepspeed} ZeRO-3~\cite{rajbhandari2021zero} model sharding. 

To understand the effect of the input language pair for the teacher scorer,
we fixed the size of the teacher
cross-encoders using the \texttt{XLM-RoBERTa-large} model with the released MS MARCO small training triples using English training and Translate-Train.
The classification head is attached to the last hidden state of the starter \texttt{<s>} token. 
We concatenate the training query and passage as the input text sequence and use cross-entropy loss to train the model.
For the English-Trained (ET) cross-encoder, we provide the native MS MARCO training queries and passages in English;
for Translate-Train (TT), we translate the passages into the document language of the final CLIR task
(Chinese, Persian, or Russian)
and use those with English queries, resulting in three TT cross-encoder models. 

\subsubsection{CLIR Student Dual-encoder.}
We use ColBERT-X, a CLIR variant of the ColBERT~\cite{colbert} retrieval architecture, as the dual-encoder model. 
Prior work has shown that single-vector dual-encoder models, such as DPR-X~\cite{c3, mmarco}, 
and learned-sparse models, such as BLADE~\cite{blade}, 
are substantially less effective than multi-vector dense dual-encoders~\cite{c3}.
Therefore, we evaluate Translate-Distill with ColBERT-X as the student model. 
Trained models are available on Huggingface Models.

We use the PLAID~\cite{plaid} implementation for ColBERT-X retrieval. 
Both evaluation queries and passages are processed in their native languages without any translation. 
The number of residual bits is set to one for each dimension of the passage representation. 
Based on our preliminary studies, the number of residual bits has a substantial impact on index size and query latency
but not on retrieval effectiveness. 
In some cases, using only one bit may result in numerically better retrieval results than using two or four bits. 

The student CLIR dual-encoder models are trained with KL-divergence loss on the predicted query-passage scores,
on a batch of 64 queries distributed across eight NVidia V100 GPUs (with the DGX platform). 
Each query is associated with six passages randomly sampled from the candidate set at every epoch.
Such random selections enable larger batches while stochastically allowing the models to see all candidate passages.
We use the AdamW optimizer with a learning rate of $5\times 10^{-6}$ with 16-bit floating point. 

Since documents in the evaluation collections are generally longer than the input length of XLM-RoBERTa, following prior work in neural IR, we create passages from the documents by using a sliding window of size 180 tokens with a stride of 90~\cite{colbertx}.
At retrieval time, we aggregate the passage scores using MaxP~\cite{maxp} to form the document score.
The number of generated passages is shown in Table~\ref{tab:collection-stats}.

\subsection{Baselines}
We compare the models trained against Translate-Train,
a predecessor to our proposed Translate-Distill training pipeline. 
We also report the effectiveness of the Patapsco~\cite{patapsco} implementation of BM25 with RM3 query expansion, using document translation into English.
This retrieval system was a baseline run for the TREC 2022 NeuCLIR track,
and was designated as the standard initial retrieval result for the reranking task~\cite{neuclir2022}. 
For the NeuCLIR collection, we use the document translations released by TREC.
We translate the HC3 documents with the same set of translation models that was used to translate the NeuCLIR collection.

\section{Results and Analysis}
\label{sec:results}

\begin{table*}[t]
\setlength\tabcolsep{0.44em}
\renewcommand{\L}{\textit{L}}
\renewcommand{\a}[0]{$^5$} %
\renewcommand{\b}[0]{$^3$} %
\newcommand{\e}[0]{$^6$} %
\newcommand{\p}[0]{\phantom{$^e$}}

\newcommand{\s}[0]{$^\dagger$} %

\newcolumntype{x}[1]{>{\centering\footnotesize\arraybackslash}p{#1}}

\caption{nDCG@20 and Judged@20 (J@20) using different teacher query-passage pair scorers.
The ``Scorer Lang.'' columns indicate the input query (Q) and passage (P) languages of the \underline{query-passage pair scorer}, respectively, where ``E'' and ``\L'' indicate English and the document language of the final CLIR task (Chinese, Persian or Russian depending on the collection), respectively.
All models other than BM25+RM3 (which indexes MT results) index documents in their native language, and all accept English queries at inference time. 
All Translate-Distill models use ColBERTv2 as the passage selector. 
The average columns report the micro-average of all 309 topics across the six collections. 
Subscript numbers indicate statistically significantly better than the system in the corresponding row with 95\% confidence using a paired $t$-test on all 309 topics. 
}

\resizebox{\linewidth}{!}{
\begin{tabular}{x{0.3cm}l|cc|ccc|ccc|l|c}
\toprule
& & \multicolumn{2}{c|}{Scorer} &  \multicolumn{7}{c|}{nDCG@20}                                   & \multicolumn{1}{c}{J@20} \\
\cmidrule{5-11}
& Query-Passage & \multicolumn{2}{c|}{Lang.} &  \multicolumn{3}{c|}{HC3} &  \multicolumn{3}{c|}{NeuCLIR 22} 
                             &  \multicolumn{1}{c|}{Micro} & \multicolumn{1}{c}{Micro} \\

& Pair Scorer           &  Q &  P &      %
\multicolumn{1}{c}{zho} & \multicolumn{1}{c}{fas} & \multicolumn{1}{c|}{rus} & 
\multicolumn{1}{c}{zho} & \multicolumn{1}{c}{fas} & \multicolumn{1}{c|}{rus} & 
\multicolumn{1}{c|}{Avg.} & \multicolumn{1}{c}{Avg.} \\

\midrule
& \multicolumn{3}{l}{\textit{Baselines}}     \\
\midrule
1 & \multicolumn{3}{l|}{BM25+RM3 using DT}
                                   &    0.262 &    0.261 &    0.136 &    0.340 &    0.355 &    0.292 &    0.301     &    0.656 \\
2 & \multicolumn{3}{l|}{BLADE}
                                   &       -- &       -- &       -- &    0.330 &    0.341 &    0.347 &    0.339\s   &    0.702\s  \\
3 & \multicolumn{3}{l|}{ColBERT-X with TT}
                                   &    0.333 &    0.411 &    0.303 &    0.441 &    0.438 &    0.470 &    0.392     &    0.547 \\ %
\midrule
& \multicolumn{11}{l}{\textit{ColBERT-X Student Models with Translate-Distill using Different Teacher Scorers}} \\
\midrule
4 & MiniLM                &  E &  E &    0.429 &    0.450 &    0.318 &    0.415 &    0.419 &    0.474 &    0.410     &    0.520 \\
5 & MonoT5-3b             &  E &  E &    0.391 &    0.510 &    0.310 &    0.479 &    0.467 &    0.497 &    0.433\b   &    0.544 \\ %
\midrule
6 & XLMR CE (ET)          &  E &  E &    0.465 &    0.501 &    0.309 &    0.452 &    0.466 &    0.503 &    0.440\b   &    0.540 \\ %
7 & XLMR CE (TT)          &  E & \L &    0.438 &    0.479 &    0.328 &    0.450 &    0.441 &    0.493 &    0.430\b   &    0.535 \\ %
\midrule
8 & Mono-mT5XXL           &  E &  E &\bf{0.473}&\bf{0.539}&\bf{0.355}&    0.492 &\bf{0.484}&\bf{0.522}&\bf{0.469}\b\a\e&    0.560 \\
9 & Mono-mT5XXL           &  E & \L &    0.468 &    0.508 &    0.332 &    0.471 &    0.475 &\bf{0.522}&    0.453\b\a &    0.548 \\
10& Mono-mT5XXL           & \L & \L &    0.453 &    0.524 &    0.342 &\bf{0.493}&    0.469 &    0.503 &    0.456\b\a\e&    0.550 \\
\bottomrule
\end{tabular}
}
{\raggedleft \s \scriptsize{Micro-average values for BLADE are computed over only the three NeuCLIR 2022 collections, and thus are not 
comparable to the other micro-averages.} }

\label{tab:main}
\end{table*}

This section presents our results for distillation, passage selection, and comparison to a Retrieve-and-Rerank pipeline.

\subsection{Distillation with a Different Query-Passage Scorer}
\label{sec:main}

We used the ColBERTv2 English passage selector for these experiments.
As shown in Table~\ref{tab:main},
ColBERT-X models trained with knowledge distilled from Mono-mT5XXL are substantially more effective than the Translate-Train baseline
(statistically significant with 95\% confidence, using paired $t$-tests and Bonferroni correction for three tests). 
The models trained with Mono-mT5XXL as the teacher scorer achieve state-of-the-art nDCG@20 performance for CLIR systems on both HC3 and NeuCLIR 2022 collections,
and they do so with a single-stage model.

As Table~\ref{tab:main} also shows,
the query-passage teacher scorer greatly affects retrieval effectiveness. 
Using publicly available English rerankers (MiniLM and MonoT5-3b) as teacher scorers does not yield more effective student dual-encoders. 
Instead, using an XLMR cross-encoder (CE) trained with English MS MARCO as the teacher scorer in English (the E-E setting),
in contrast to using it as a reranker in a Retrieve-and-Rerank pipeline~\cite{monobert},
provides better supervision for training the student dual-encoders.
While the difference between XLMR CE with ET and MonoT5-3b is not statistically significant, the fact that an English-trained XLMR cross-encoder distills to the student ColBERT-X model as well as the MonoT5-3b suggests that the model size of the teacher scorer is not a dominant factor in the training pipeline. 
Among all teacher scorers operating in monolingual English, 
using Mono-mT5XXL as the teacher scorer results in statistically significantly more effective student ColBERT-X models (Row 8) %
than using either the MonoT5-3b teacher (Row 5) %
or the XLMR English-trained CE (Row 6) %
We hypothesize that the Mono-mT5XXL model is particularly well-trained and in this case the additional power does not come from the model size. 
However, we cannot eliminate the possibility that
growing the model size even further would eventually lead to improvements based solely on size. 
We leave this hypothesis for future investigation.

The languages of the input query and passages for the teacher models also substantially affect retrieval effectiveness. 
For both XLMR cross-encoder and MonoT5XXL as the teacher scorer,
training student models with teacher scorers operating in the language of the training corpus (English in this case) is more effective than using document languages. 
Keeping the input MS MARCO query-passage pairs for the teacher scorer in English (the E-E setting)
when using the English-trained (ET) XLMR cross-encoder (CE) produces a more effective subsequent ColBERT-X model
than when using its Translate-Trained (TT) counterparts (the E-L setting).
Using MonoT5XXL as the teacher scorer shows a similar trend. 
Such results indicate that the teacher scorer should be trained using the original text without translation (English for MS MARCO in this case).

Nevertheless, these observations suggest that language knowledge can be passed down through the teacher scores for MS MARCO training queries and passages. 
While the \textit{vessels} for the knowledge transfer (the training queries and passages) are the same
(because the same passage selector was used),
the \textit{cargo} carried by those vessels (the scores) differ;
this can have a large effect on what the student model learns. 
In the next section, we explore the effects of changing the \textit{vessel}.

\subsection{Passage Selector}
\begin{table*}[t]
\setlength\tabcolsep{0.45em}
\renewcommand{\L}{\textit{L}}

\caption{nDCG@20 and R@1000 on NeuCLIR 22 Chinese using different passage selectors and query-passage pair scorers. The tested passage selectors include English ColBERTv2~(CBv2), English CoCondenser~(CoC), and English-Chinese Translate-Trained ColBERT-X~(CB-TT).}
\centering

\newcolumntype{x}[1]{>{\centering\arraybackslash\hspace{0pt}}p{#1}}

\begin{tabular}{lx{0.65cm}x{0.65cm}|ccc|ccc}
\toprule
& \multicolumn{2}{c|}{Scorer Lang.}  & \multicolumn{3}{c|}{nDCG@20} & \multicolumn{3}{c}{R@1000}\\

Pair Scorer &   Q &   P &      CBv2 &     CoC &   CB-TT &      CBv2 &     CoC &       CB-TT \\
\midrule
MiniLM      &   E &   E & \bf{0.415}&   0.408 &   0.397 &     0.840 &     0.819 & \bf{0.853}\\

Mono-mT5XXL &  \L &  \L & \bf{0.493}&   0.483 &   0.460 &     0.867 &     0.867 & \bf{0.884}\\
\midrule
Average     &     &     & \bf{0.454}&   0.446 &   0.429 &     0.853 &     0.843 & \bf{0.869}\\
\bottomrule

\end{tabular}

\label{tab:diff-selector}
\end{table*}

The previous section focused on the effect of the query-passage pair scorers when using ColBERTv2 as the passage selector.
The following experiments vary the passage selector, using Chinese as an example.
Table~\ref{tab:diff-selector} shows the results of those experiments.
There was no statistically significant difference between the two English selectors (ColBERTv2 and CoCondenser) for nDCG@20 or Recall@1000 (R@1000),
although models trained with passages selected by the ColBERTv2 model achieved numerically higher results by both measures.
Interestingly, using a Translate-Trained ColBERT-X as the passage selector does not lead to a more effective student model when measured by nDCG@20,
but it does result in a more effective student model when measured by R@1000. 
We hypothesize that the Translate-Trained ColBERT-X model is not effective enough to retrieve sufficiently hard (and thus informative) candidate passages. 
Since it selects passages using their translated text,
the candidate sets contain a larger diversity in the document language (Chinese, in this set of experiments),
resulting in student models with higher Recall@1000. 

Given these observations, we conclude that both the query-passage pair scorer and the passage selector affect final CLIR retrieval effectiveness.
The recipe for creating an effective CLIR dual-encoder model requires that the selected passages are sufficiently hard,
for which we can use monolingual English neural retrieval models. 
Also, the subsequent query-passage pair scorer should be selected carefully
to be both effective and aligned with the native language of the training corpus (English for the MS MARCO).

\subsection{Comparison to Retrieve-and-Rerank Pipeline}

Finally, we compare two ways of using a cross-encoder: reranking and distillation. 
To facilitate our comparison, the cross-encoder we use as teacher for our student dual-encoders is also used as the reranker of the top 200 first-stage results in the Retrieve-and-Rerank pipeline (R\&R).
To eliminate the effect of translation and potential translationese
(language artifacts from the machine translation models),
none of our retrieval pipelines uses query or document translation.
Instead, we use probabilistic structured queries (PSQ) with HMM~\cite{darwish2003probabilistic, wang2012matching} as the first-stage retriever,
which is a strong sparse CLIR baseline model that does not require one-best neural machine translation. 
PSQ indexes documents in the query language tokens (English in our experiments)
by translating the document tokens probabilistically using an alignment matrix. 
To be more specific, each token in the original document is translated into a bag of query language tokens,
where the weight of each resulting token is the product of the original weight multiplied by the translation probability. 
Therefore, each translated document is a bag-of-token with probabilistic weights in the query language that can be indexed by sparse retrieval systems such as Lucene. 

The rerankers are also restricted to those capable of accepting queries and documents in different languages. 
Note that the official TREC 2022 NeuCLIR Track first-stage retrieval results used BM25 with document translation,
which differs from the setting here.
However, to compare systems under identical conditions, we use the PSQ model in the pipeline,
which is a slightly weaker but much more efficient CLIR system~\cite{blade}. 

As summarized in Table~\ref{tab:rerank}, models trained with Translate-Distill provide at least 94\% of their teachers' retrieval effectiveness. 
For XLMR cross-encoders, the distilled ColBERT-X models even outperform the use of their cross-encoder teacher models as rerankers. 
While a Retrieve-and-Rerank pipeline using Mono-mT5XXL is no better than student dual-encoders trained with its knowledge,
using the cross-encoders for reranking is much more computationally expensive at retrieval time,
and thus query latency could be much longer than for an end-to-end ColBERT-X model. 

\begin{table*}[t]
\setlength\tabcolsep{0.5em}
\renewcommand{\L}{\textit{L}}
\renewcommand{\delta}{$\triangle$}

\caption{nDCG@20 on NeuCLIR 2022 without query or document translation in the retrieval pipeline. R\&R indicates the retrieve-and-rerank pipeline using PSQ as the first-stage retriever followed by the cross-language cross-encoder reranker specified in the first column. T-D indicates the score of the ColBERT-X model trained with Translate-Distill using the cross-encoder specified in the first column as the teacher scorer. }
\centering

\begin{tabular}{l|rrr|rrr|rrr}
\toprule
                         &       \multicolumn{3}{c|}{zho} &       \multicolumn{3}{c|}{fas} &        \multicolumn{3}{c}{rus} \\
{}                       &     R\&R &      T-D &   \delta &     R\&R &      T-D &   \delta &     R\&R &      T-D &   \delta \\
\midrule
\textit{First Stage PSQ} &    0.329 &       -- &       -- &    0.358 &       -- &       -- &    0.330 &       -- &       -- \\
\midrule
XLMR CE (ET)             &    0.299 &\bf{0.452}&    151\% &    0.306 &\bf{0.466}&    152\% &    0.355 &\bf{0.503}&    142\% \\
XLMR CE (TT)             &    0.371 &\bf{0.450}&    121\% &    0.362 &\bf{0.441}&    122\% &    0.405 &\bf{0.493}&    122\% \\
Mono-mT5XXL              &    0.459 &\bf{0.493}&    107\% &\bf{0.501}&    0.469 &     94\% &    \bf{0.503} &\bf{0.503}&    100\% \\
\bottomrule
\end{tabular}

\label{tab:rerank}

\end{table*}
\section{Conclusions and Future Work}
This work introduces Translate-Distill, a training pipeline that distills ranking knowledge from an effective cross-encoder to train a CLIR student dual-encoder model. 
Evaluated on two CLIR collections, each with three language pairs,
we have shown that this training pipeline produces statistically significantly more effective CLIR dual-encoders than the earlier Translate-Train approach. 
Dual-encoders trained with Mono-mT5XXL as the query-passage pair scorer achieve state-of-the-art effectiveness for CLIR end-to-end neural retrieval on the TREC 2022 NeuCLIR benchmark.  

We expect that Translate-Distill may help not just with dual-encoders,
but also with other neural retrieval models. 
For example, work on BLADE~\cite{blade}
(a CLIR variant of SPLADE~\cite{formal2021splade})
has shown Translate-Train to be beneficial;
perhaps it may also benefit from Translate-Distill. 
Distilling from even larger models such as GPT-4, which is known to be a multilingual model~\cite{jiao2023chatgpt}, may further improve the student CLIR models.

\bibliographystyle{splncs04nat}
\bibliography{bibio}

\end{document}